# A Gating Grid Driver for Time Projection Chambers

S. Tangwancharoen[1,2], W.G. Lynch[1,2], J. Barney[1,2], J. Estee[1,2], R. Shane[1], M.B. Tsang[1,2*], Y. Zhang[3], T. Isobe[4], M. Kurata-Nishimura[4], T. Murakami[5], Z.G. Xiao[3], Y.F. Zhang[6] and the SπRIT collaboration

*corresponding author: tsang@nscl.msu.edu

[1]*National Superconducting Cyclotron Laboratory, Michigan State University, East Lansing, MI 48824, USA*
[2]*Department of Physics and Astronomy, Michigan State University, East Lansing, MI 48824, USA*
[3]*Department of Physics, Tsinghua University, Beijing 100084, China*
[4]*RIKEN Nishina Center, Hirosawa 2-1, Wako, Saitama 351-0198, Japan*
[5]*Department of Physics, Kyoto University, Kita-shirakawa, Kyoto 606-8502, Japan*
[6]*College of Nuclear Science and Technology, Beijing Normal University, Beijing 100875, China*

## ABSTRACT

A simple but novel driver system has been developed to operate the wire gating grid of a Time Projection Chamber (TPC). This system connects the wires of the gating grid to its driver via low impedance transmission lines. When the gating grid is open, all wires have the same voltage allowing drift electrons, produced by the ionization of the detector gas molecules, to pass through to the anode wires. When the grid is closed, the wires have alternating higher and lower voltages causing the drift electrons to terminate at the more positive wires. Rapid opening of the gating grid with low pickup noise is achieved by quickly shorting the positive and negative wires to attain the average bias potential with N-type and P-type MOSFET switches. The circuit analysis and simulation software SPICE shows that the driver restores the gating grid voltage to 90% of the opening voltage in less than 0.20 $\mu$s. When tested in the experimental environment of a time projection chamber larger termination resistors were chosen so that the driver opens the gating grid in 0.35 $\mu$s. In each case, opening time is basically characterized by the RC constant given by the resistance of the switches and terminating resistors and the capacitance of the gating grid and its transmission line. By adding a second pair of N-type and P-type MOSFET switches, the gating grid is closed by restoring 99% of the original charges to the wires within 3 $\mu$s.

I. **Introduction**

Since its invention **[1]**, Time Projection Chambers (TPCs) have been used successfully in many experiments to measure charged particles emitted in nuclear collisions, using devices such as the EOS TPC **[2] [3]**, the CERES/NA45 Radial Drift TPC **[4]**, the NA49 large acceptance hadron detector **[5]**, the STAR detector at Relativistic Heavy Ion Collider (RHIC) **[6]** and the ALICE detector at the Large Hadron Collider (LHC) **[7, 8]**. In this article, we will use the SAMURAI Pion-Reconstruction Ion-Tracker Time Projection Chamber (SπRIT-TPC) **[9]**, designed for use with the SAMURAI spectrometer at the Radioactive Ion Beam Factory (RIBF) at RIKEN, Japan **[10]** to illustrate the properties of the new gating grid driver.

The operation principle of a TPC and its wire planes are illustrated in Figures 1 and 2. Figure 1 shows a TPC field cage, which is filled with counter gas. Electrodes on the walls of the field cage provide a uniform electric gradient potential within the cage. The TPC is normally placed inside a uniform magnetic field, which is anti-parallel to the electric field. The magnetic field allows the determination of the momenta of charged particles and has the ancillary benefit of improving the resolutions of particle tracks by limiting the diffusion of drift electrons in directions perpendicular to the magnetic field.

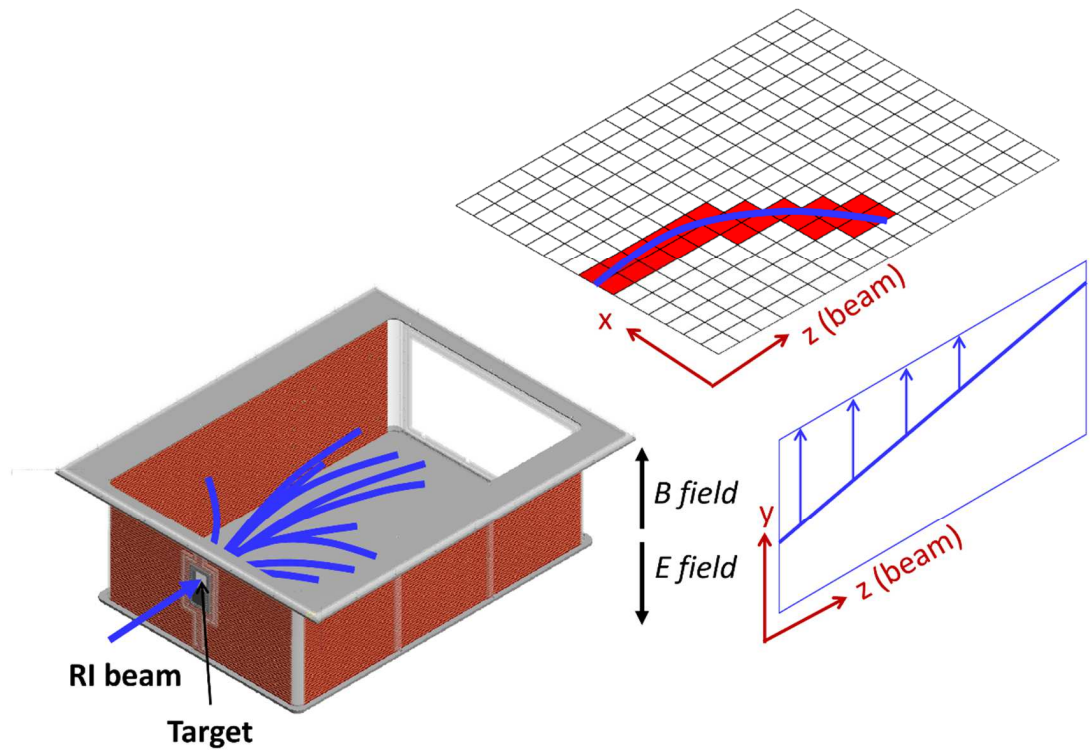

**Figure 1:** *Schematic representation of tracks of charged particles in the field cage of a TPC. The field cage provides a uniform electric field, **E**. Momenta of the particles can be obtained by placing the TPC inside a uniform magnetic field, **B** that is anti-parallel to **E**. By requiring uniform and anti-parallel fields **E** and **B** fields, one minimizes **E** × **B** forces on the drift electrons and improves the reconstruction of the tracks.*

When a violent heavy-ion reaction occurs, fast charged particles are produced in the target, which is located just outside of the upstream window of the field cage. These charged particles enter the field cage through the window and ionize the detector gas, liberating electrons that will be referred to as drift electrons. These drift electrons move along the anti-parallel electric and magnetic fields towards a set of three wire planes located at the top of the field cage. The wire planes are not visible in Figure 1, but their functions are illustrated schematically in Figure 2, in which the wires are drawn larger than scale to make them more visible. The 3 layers of wire planes are mounted just below a pad plane tiled with pads. This pad plane forms the upper boundary of the field cage volume, and the bottom boundary of the field cage is the cathode. By measuring the arrival time and the induced charge of the avalanche electrons produced around the anode wires, the TPC provides an accurate 3-D reconstruction of these tracks in the gas, from which the particle momenta and the energy loss of each charged particle detected in the counter gas can be deduced.

In many TPC applications, there are charged particles that enter the field cage that are not of scientific interest. In the SπRIT TPC experiments **[9, 11]**, these include beam particles that do not interact with the target or large projectile residues from very peripheral collisions. It is important to prevent the gas multiplication of the drift electrons from such undesired particles. Amplification of undesired events will accelerate the aging process of the anode wires by creating negatively charged polymers from the hydrocarbon components or impurities in the detector gas. If the deposition of such polymers on the anode wires is not controlled, the effective anode wire diameters can increase with time due to the deposit, reducing the gas gain and deteriorating the performance of the TPC **[12]**.

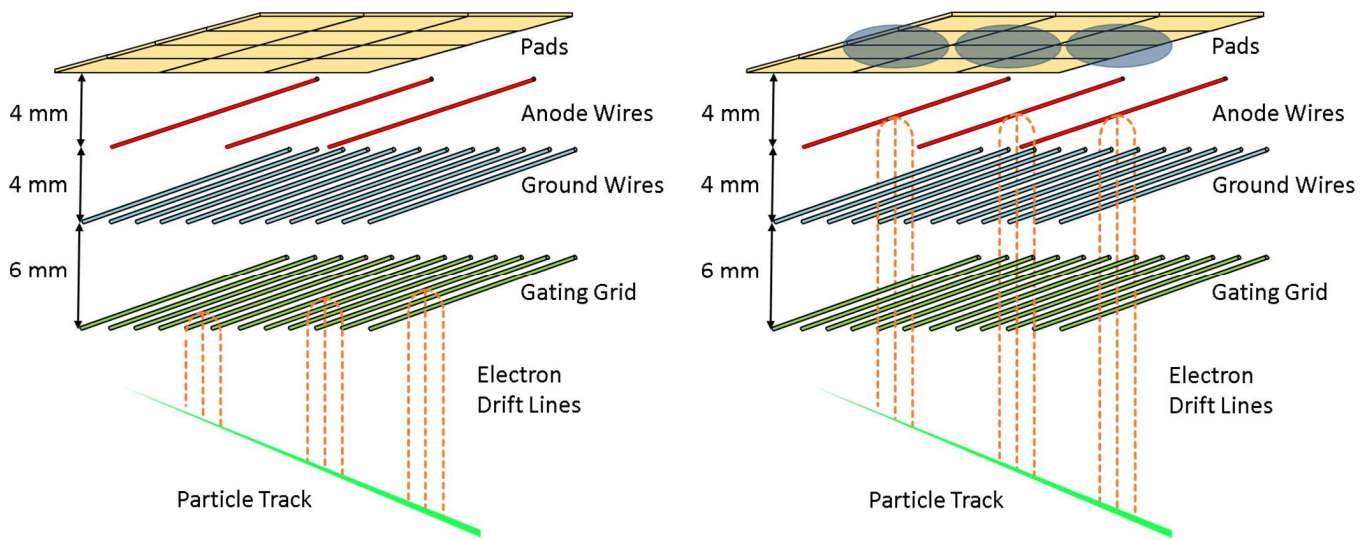

**Figure 2:** *A cartoon illustrating the closed (left panel) and open (right panel) state of the gating grid. For ease of viewing, the wire diameters and heights shown are not to scale. The wires run parallel to the x-axis. The z-axis defined by the beam in the direction of the particle track is orthogonal to the wires. The drift time of the electrons to the anode plane provides the vertical (y) location of the ionization track and the induced charge of the electrons at the pad plane provides the horizontal locations of the ionization track.*

To suppress the detection of unwanted particles, it is essential that the wire plane closest to the drift region, called the gating grid, remains "*closed*" to drift electrons resulting from the unreacted beam and other *uninteresting* events. In this closed state, as illustrated in the left panel of Figure 3a, the gating grid captures the drift electrons produced in the field cage volume. When the external trigger detection system indicates the occurrence of an *interesting* event, the gating grid is *opened* as illustrated in Figure 3b (right panel). Then, the drift electrons pass through the gating grid, and then through the ground wire plane, to reach the anode wires located between the ground plane and the pad plane. These drift electrons will then trigger an avalanche in the high electric field region of the anode wires, which multiplies the electrons by a typical gas gain of about 2000 depending on the anode voltage.

This avalanche also produces positive ions, whose motions away from the anode wires generates image currents on the pad plane as illustrated in Figure 2b. These currents are amplified by the TPC electronics located above the pad plane and are recorded. After their production, many of these positive ions travel through the ground plane towards the gating grid and the drift region of the TPC field cage. The drift velocities (on the order of cm/ms) of these positive ions are much slower than those of the electrons. It is important that the gating grid is closed to prevent these positive ions from passing through and accumulating in the drift region of the TPC field cage. Otherwise, the space charge of these positive ions could seriously distort the electric field of the field cage and degrade the reconstruction of the tracks of the reaction products. To avoid this, the gating grid remains closed after all the drift electrons from an interesting event pass through it and only opens when triggered by the next interesting event.

II. **Gating Grid**

The electrostatic potentials on the gating grid wires are used to control the passage of drift electrons and positive ions and allow one to separate the drift volume of the TPC from the avalanche region. The gating grid serves three functions:

1. It prevents drift electrons from unwanted events from going into the avalanche region and being multiplied
2. It prevents the back flow of the positive ions from the avalanche region into the drift volume.
3. It minimizes damage to the anode wires caused by the deposition of ionized polymers.

The gating grid described here operates in a "bipolar" mode. When the gating grid is open, the potentials of all the wires are set to a common voltage, $V_a$, to match the electric field in the field cage and maximize the transmission of drift electrons. This makes the wires almost invisible to the drifting electrons, which pass through the gating grid plane and ground wire plane to the avalanche region. To close the gating grid in the bipolar mode, alternate wires are biased to potentials of $V_h = V_a + \Delta V$ and $V_l = V_a - \Delta V$. The gating grid has positive surface charge density on every other wire and negative surface charge density on the wires in between. When the voltages are suitably chosen so as to fully close the gating grid, the electron drift lines will terminate at the positive wires (connected to $V_h$), and the positive ions will terminate on the negative wires (connected to $V_l$). This effectively closes the gate and the drift electrons and positive ions cannot pass through. The advantage of the bipolar mode is that the potential difference between open and closed configuration is very small at the location of the pad plane. Thus the pickup noise charge induced on the pads during the transition between the open and closed configurations of the gating grid can potentially be small.

For illustration, we have modeled the electric field and the electron drift in the SπRIT TPC with Version 9 of the Garfield drift chamber simulation program [13] and its associated toolkit of programs, such as Magboltz [14]. The SπRIT TPC wire plane configuration is given in Table 1 and illustrated in Figure 2.

In the simulations shown in Fig. 3 and 4, we assumed P10 gas, consisting of 90% Argon and 10% $CH_4$ by volume as the counter gas at a pressure of 101.3 kPa. The voltages of the gating grid were set to $V_a = -110$ V and $\Delta V = 70$ V. Simulations were performed for a drift field of 131 V/cm in the field cage, corresponding to the maximum in the electron drift velocity for P10 gas at zero magnetic field.

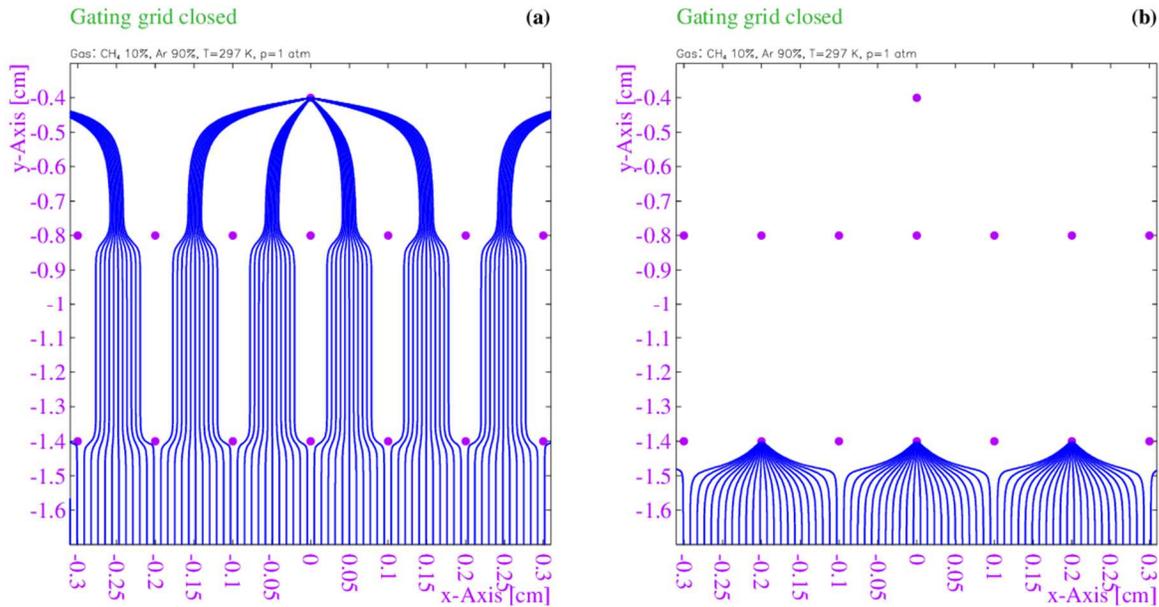

**Figure 3:** *Expanded view of electron drift lines near the wire planes for the configuration a. (Left panel) with the gating grid open and b. (Right panel) with the gating grid closed. The gating grid, ground and anode wires are located at y=−1.4 cm, -0.8 cm and -0.4 cm respectively.*

Figure 3a, (left panel) shows the calculated electric drift lines near the wire planes when the gating grid is open. All electrons pass from the drift volume and continue between the wires of the gating grid and ground plane, terminating on the anode wire where avalanche occurs. In the coordinate system used in the figure, y=0 cm corresponds to the pad plane. For simplicity, Figure 3a shows only a subsection of the wire planes containing one anode wire, which is located at y = −0.4 cm and x = 0 cm. Similarly, we show 7 ground plane wires, located at y = −0.8 cm (x = 0 to 0.8 cm) and 7 gating grid wires, located at y = − 1.4 cm (x = 0 to 0.8 cm). The wires in both ground and gating grid planes are separated by a pitch of 1 mm as listed in Table 1.

| Wire-plane specifications | | | | | |
|---|---|---|---|---|---|
| Plane | Material | Diameter (μm) | Pitch (mm) | Distance to pad plane (mm) | No. of wires |
| Anode | Au-plated W | 20 | 4 | 4 | 364 |
| Ground | BeCu | 76 | 1 | 8 | 1456 |
| Gating | BeCu | 76 | 1 | 14 | 1456 |

**Table 1:** *Wire–plane specifications for the SπRIT Time Projection Chamber*

The gating grid is closed by biasing the potentials on neighboring wires alternatively. Figure 3b (right panel) shows the electric drift lines when neighboring wires are biased to −180 V and −40 V, attracting electrons to the more positive wires at −40V and repelling them from the neighboring more negative wires biased to −180 V. Neglecting scattering in the gas, nearly all electrons from the drift volume will follow the electric field drift lines, terminating at the nearest adjacent positively biased wire, effectively preventing significant transport of electrons through the gating grid to the anode wires. Small leakage may occur when electron-ion scattering is taken into account. This is typically counteracted by increasing the difference voltage **ΔV**.

Many positive ions are produced by the electron avalanche at the anode wires. The ions are repelled from the anode wires inducing an image of the avalanche in the pad plane. They are also attracted to the ground wire where most of them terminate. Any positive ions that pass through the ground wires will be attracted to the more negative grid wires biased to −180 V and will be repelled by the neighboring wires. Nearly all ions should be stopped by the ground and gating grid wires.

The behavior of drifting electrons and ions is strongly influenced by the magnetic field. Its effect can be characterized by the parameter $\omega\tau$ where $\omega$ is a cyclotron frequency of the charged particle, $\omega = qB/m$, (electron or ion), $\tau$ is the mean time between collisions of the charged particle with the gas molecules, q is the charge and m is the mass of the particle. Due to the large masses, $\omega\tau$ is very small $O(10^{-4})$ for ions, but can be greater than unity for electrons in certain gases and magnetic fields. Larger corrections occur for the electrons in the parts of their trajectories where the electric field from a neighboring wire is perpendicular to magnetic field. The magnetic field reduces the mobility of electrons in directions towards the gating grid wires while adding components to the electron velocities in a direction parallel to the wire. Both magnetic effects diminish the number of drift electrons captured on the wire for a set of electrostatic potential when one increases the magnetic field, thereby increasing the difference voltage **ΔV** required to close the gate **[15]**.

Figure 4 shows the electron transparency of a gating grid in the presence of magnetic field. The electric field ***E*** is anti-parallel to the magnetic field ***B***. The voltage required to close the gate increases with the magnetic field, from **ΔV** ≥ 25 V without magnetic field to **ΔV** ≥ 50 V with magnetic field of 0.5 T and **ΔV** ≥ 70 V with magnetic field of 1 T. To minimize the statistical uncertainties in the simulations, 4000 electron trajectories were sampled in each set of calculations.

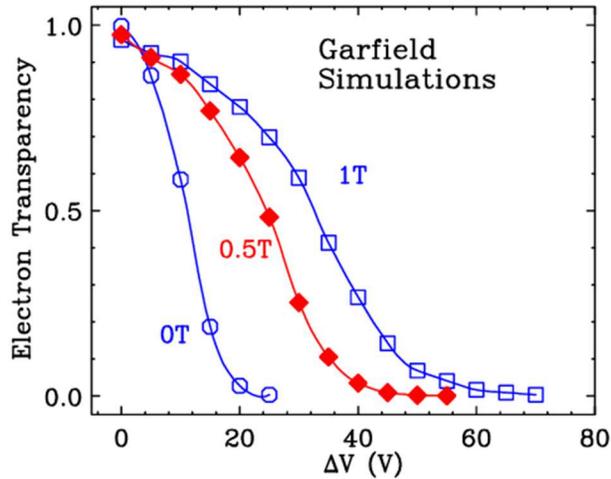

**Figure 4:** *Electron transparency as a function of ΔV from Garfield simulations.*

### III.     Gating Grid Driver

The gating grid driver controls the transition between the closed and open states of the gating grid. After the event trigger is satisfied, drift electrons will continue to drift into the grid and be captured, unless the driver opens the gating grid fully. The time elapsed between the arrival of the trigger and opening the gate fully creates a "dead" region immediately below the gating grid of thickness $\Delta L_{dead} = v_{drift} \cdot \Delta t_{open}$. Drift electrons produced in this region will be lost. To minimize $\Delta L_{dead}$, it is clearly important to design the driver to open the gating grid as quickly as possible. For P10 gas at atmospheric pressure with a magnetic field of 0.5 T, the typical $v_{drift} \approx 5.45$ cm/μs **[6]**. With these operating conditions and an opening time of 0.4 μs, the dead region is 2.2 cm thick.

When the gating grid is closed, alternate the wires are biased to $\mathbf{V_h} = \mathbf{V_a} + \Delta \mathbf{V}$ and $\mathbf{V_l} = \mathbf{V_a} - \Delta \mathbf{V}$. To open the gating grid, the voltages on half of the wires must be reduced by $\Delta \mathbf{V}$ and the voltage on the remaining wires increased by $\Delta \mathbf{V}$. This can potentially induce a large signal on the pads that could be comparable to the signal induced by a weakly ionizing particle such as a pion. To minimize such effects, a well-designed gating grid driver should change these charges at equal rates from both positive and negative sides of the wires so that the average potential of the gating grid remains constant minimizing the induced signal (noise) on the pads. The closure of the gating grid is achieved by restoring the original alternating potentials.

#### A. Design of gating grid driver circuits

A conventional gating grid driver employs three high voltage power supplies, one at the common voltage ($\mathbf{V_a}$), one at the higher voltage ($\mathbf{V_h}$) and the third at lower voltage ($\mathbf{V_l}$) **[16]**. The switching of voltages is typically accomplished with 4 switches, one pair for each alternate set of wires that switches the potential from the "closed" voltage ($\mathbf{V_h}$ or $\mathbf{V_l}$) to the average "open" voltage ($\mathbf{V_a}$) and back to the "closed" voltage again.

In our design, the common voltage $\mathbf{V_a}$ is set by shorting the more positive and more negative gating grid wires together to open the gate. This opening current flow differs from the conventional design, where both positive and negative wires are separately connected to $V_a$, especially if the impedances of the connections to $V_a$ for positive lines and negative lines differ. Later, after the electrons from the drift region are collected, the gating grid is closed in our design by connecting the positive and negative wires back to their original voltage supplies at $V_h$ and $V_l$, respectively. The resulting current flows in our current design during closing are rather similar to that of the conventional design. The key challenge in any gating grid is to open the grid rapidly without having the average voltage of the two sides of the gating grid deviating significantly from the correct average value $V_a$. We designed the driver circuit to minimize the opening time, corresponding to 90% transparency in about 350 ns in experimental conditions. Since the subsequent readout dead time negates any urgency for closing the gating grid quickly, we chose to reduce the currents by lengthening the gating grid closing time to about 3 μs.

## B. MOSFET Switches

The gating grid driver uses two pairs of N-MOSFET and P-MOSFET switches. The N- and P- type MOSFET switches work in tandem and are driven by MOSFET drivers, which are controlled by TTL signals. The advantage of using a MOSFET switch is that when the switch is closed, the internal resistance is lower compared to other types of switches. This helps in allowing one to tune the resistance across the switches to a lower total resistance value. Also, a MOSFET switch has a short turn-on delay time, which is the time taken to charge the input capacitance of the device before draining current conduction can start. This allows the gating grid to be opened quickly. Since the source side of the MOSFET switch defines the reference voltage of the gate signal to the switch, we connect the negative side of the gating grid to the drain connection of a P-MOSFET [16] and the positive side of the gating grid to an N-MOSFET switch [17]. Both P-MOSFET and N-MOSFET switches have the same turn-on delay time of 14 ns.

To drain charges from the gating grid as fast as possible, we drive the MOSFET switches at the saturated region where the internal resistances of the switches achieve their smallest values of 0.18 Ω for IRF640 and 0.5 Ω for IRF9640. These MOSFET switches are driven by the gate driver chips that supply the gate signal to control the operation of N- and P- switches. To achieve the saturated region for these MOSFET switches, the voltages difference between the gate and source terminals should be greater than or equal to 10 V for N-MOSFET and less than or equal to -10 V for P- MOSFET switches. To provide this voltage, we use a MIC4420 chip for each N- MOSFET switch and a MIC4429 chip for each P- MOSFET switch. Powered by a 12 V power supply and controlled by TTL signals, these chips can provide the required voltage difference between gate and source terminals from 0 V to the operating voltage of ±12 V in 20 ns, safely below the maximum voltage of ±18 V. Unlike the MOSFET switches, which are floating at the average voltage, the grounds of the MIC4420 and MIC4429 chips are referenced to the external power supplies, and the outputs of these gate drivers are capacitively connected to the MOSFET switches. In addition to these switches that combine to open the gating grid, there are two switches of the same type that are used to close the gating grid in about 3 µs by recharging the gating grid to its original voltages of $V_h$ and $V_l$. Thus, four switches, two HV power supplies and one 12 V supply are required to open and close the gating grid.

### C. Some differences with prior gating grid driver designs

Ref. **[16]** describes the ALICE gating grids, which have a capacitance of $C_{grid}$ = 6 nF. They are connected to the gating grid driver via standard $R_{line}$ = 50 ohm transmission lines. Upon initiating the opening of the gating grid, the ALICE voltage difference $\Delta V$ decays exponentially with a decay constant $\tau = R_{line} \cdot C_{grid}$ = 300 ns. This decay constant means that $\Delta V$ decreases to 10% of its original value in about $\Delta t$ = 690 ns.

With the same driver and cable impedance, our gating grid with its larger capacitance of about 16 nF, would drop to 10% of its original value in $\Delta t$ = 1.84 µs corresponding to a lost drift length of $\Delta L_{dead}$ = 10 cm. For the SπRIT TPC, such an opening time is unacceptably long because the drift distance between the beam line and gating grid is only 19 cm. To reduce this opening time, the present gating grid driver circuit uses 4 ohm transmission lines to reduce the opening time below that achievable with conventional 50 ohm transmission lines. It should be noted that that capacitance of the transmission line adds about 4 nF/m of capacitance to the overall capacitance. It is therefore advantageous to adopt a short cable between the driver and TPC. The short cable length means that actual decay constant is mainly governed by the terminating resistances in the gating grid driver, which could be varied by a factor of two from their impedance matched values without causing large reflections. Geometric details of such internal connections are very specific to the TPC design so we do not explore them further here.

### D. Operation of the gating grid driver

Figure 5 shows the circuit diagram of the gating grid driver. To initiate the opening of the gating grid, one applies a positive TTL signal to the input connection labeled TTL1 which closes the MOSFET switches labeled N1 and P1 on the right edge of Circuit A. Once these switches are closed, the current flows between the two sides of the gating grid until both sides reach the common voltage. The length of the TTL1 signal defines the length of the time that these switches remain closed. The gating grid should remain open long enough to let all the drift electrons from the interesting event to pass through the grid.

To close the gate, one ends TTL1 signal and promptly afterwards sends a TTL signal to the TTL2 input, which then closes the two switches labeled N2 and P2. The two sides of the gating grid are connected to two high voltage supplies, labeled HV-High and HV- Low, which charge the high and low voltage sides to $V_h$ and $V_l$, respectively, through the two 10 Ω resistors shown in the figure. This is accomplished in approximately 3 $\mu$s. After 3 $\mu$s has elapsed, the N2 and P2 switches open and the voltage on the gating grid is maintained by two 18 kΩ resistors that connect the positive and negative sides of the gating grid to the HV supplies.

To avoid inducing unwanted signals on the pads when the gating grid is initially opened, the positive and negative wires of the grid must change their voltages at rates that are equal in magnitude and opposite in sign. This rate is given by the RC time constant $\tau$ of circuit A, where R≈Rp+Rn and C is the capacitance of the gating grid and cable network. More detailed analysis of this time dependence is given in the next section.

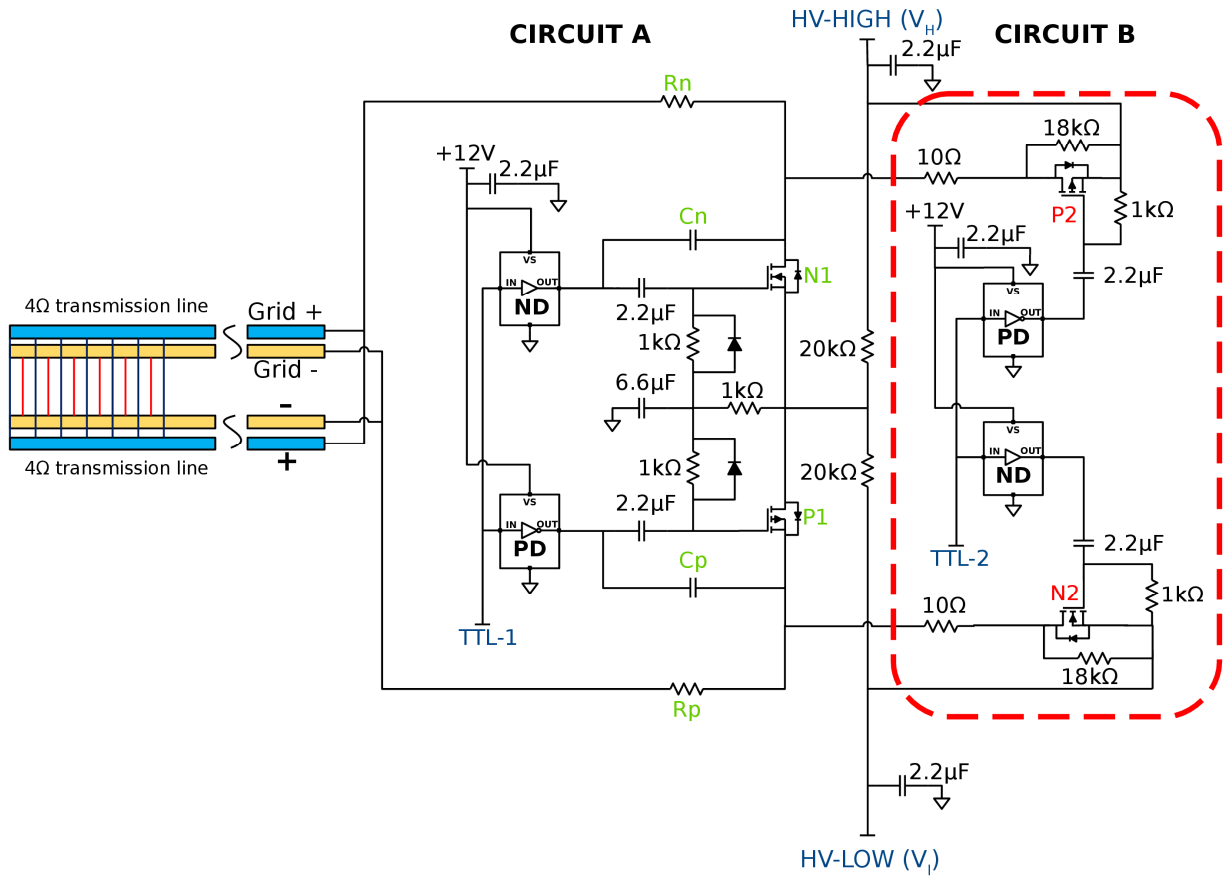

**Figure 5:** *Circuit diagram of the gating grid driver. Circuit B refers to the circuit inside the red-dashed area, and circuit A refers to the rest of the circuit.*

**SPICE: circuit analysis program**

Throughout this study, we use the circuit analysis program, SPICE **[19]**, to simulate the gating grid driver in order to understand the properties of the driver circuit, problems that we encounter and also to provide guidance how to tune the gating grid driver. SPICE enables DC, AC and transient analysis and can be used to check the integrity of the circuit designs and to predict the performance of a circuit. Due to its popularity, many manufacturers provide SPICE models for their electronic components to facilitate the simulation of the performance circuits made with these components. The simulations described below, were performed with the version OrCAD EE PSPICE downloaded from **[20]**.

To study the discharge characteristics of the MOSFET switches in our circuit, we use the SPICE model of the MOSFET switches provided by Vishay **[17, 18]**. In the simulation, the gating grid is modeled by a capacitor of 26.5 nF, which matches the capacitance of the S$\pi$RIT TPC gating grid **[9]** in its operating environment, including cables between TPC and driver. Figure 6 shows SPICE simulations of the gating grid switching from the closed configuration to the open configuration and then back to closed configuration. At the beginning of the simulations, the gating grid is closed. Thus, the alternate wires in the gating grid are biased to −40V and −180 V. A TTL1 logic signal is input to the circuit at t=0.5 $\mu$s, initiating the closing of the N- and P- MOSFET switches. It takes an additional 0.05$\mu$s for the N1 and P1 switches to open and the voltages to begin their exponential decrease. In this circuit, there are two resistors Rp and Rn and two capacitors Cp and Cn that can be used to tune the performance of the gating grid driver.

Figure 6 shows the decrease in voltage of the gating grid for default values of Cp=Cn=100 pF and Rp=0.95 $\Omega$, Rn=1.05 $\Omega$. With these values, $V_h$ (red) and $V_l$ (blue) are within 10V of a common voltage of −110 V, or in effect 90% open, 0.20 $\mu$s after switches N1 and P1 open (left panel, Fig. 6a). Their average voltages are shown in black. In this particular case, TTL1 is 4 $\mu$s long. At the end of the TTL1 signal, the TTL2 signal arrives and the gating grid begins to close. The gating grid wires are restored to 99% of their original voltages within 3 $\mu$s as illustrated in Figure 6b (right panel).

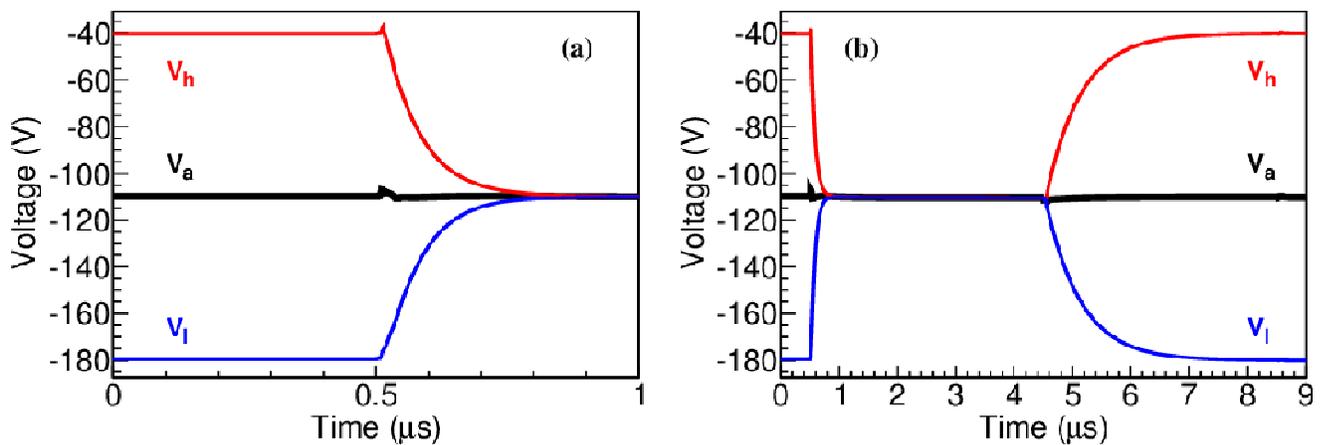

**Figure 6:** *(a) Transition of the gating grid from closed to open and (b) from closed to open and back to closed state in a SPICE simulation. Cp=100 pF, Cn=100 pF and Rp=0.95 $\Omega$, Rn=1.05 $\Omega$.*

Figure 6 shows the results of SPICE simulation of an ideal circuit with ideal switches. In reality, both switches and other circuit components may have properties that are poorly documented. More importantly, the gating grid in the environment of the TPC will be capacitively and inductively coupled to the ground plane, the anode plane and the pad plane. These couplings are specific to the design of the gating grid. Thus, we include a pair of adjusting capacitors, Cp, Cn and a pair of terminating resistors Rn and Rp associated with the N1 and P1 switches to allow some compensation of couplings that appear in the real environment of a TPC gating grid. Typically, Rn and Rp control the discharge and balance the discharge rates on the positive and negative side of the gating grid, while Cn and Cp can be adjusted to separately balance the positive and negative charge.

By changing the values of Cp, Cn, Rp and Rn, one can explore situations when the discharge rates of the positive and negative sides of the gating grid are not the same. Figure 7a shows the calculated voltages on the more positive (red) and negative (blue) sides of the gating grid and their average (black) on an expanded scale that focuses on the opening of the gating grid. These calculations were performed for Cp=600 pF, Cn=100 pF and Rp=0.95 Ω, Rn=1.05 Ω. For illustration, we choose Cp >>Cn to represent some aspect of the fabrication in which one side of the gating grid has a stronger capacitive coupling to the ground plane or to the TPC ground than does the other side, which might occur due to some space requirement imposed by the geometry of the TPC or its cabling. The average (black) signal shows that this asymmetry introduces a significant asymmetry in the average voltage obtained by adding the positive and negative voltages. This average voltage reaches an extremum, $V_X = -122.3$ V, at point X, which is 12.3 V lower than the base average voltage of $V_a = -110$ V. As the gating grid opens, the difference between the voltages on the two sides of the gating grid vanishes. The point labeled B in the figure corresponds to the point that $\Delta V \leq 10$ V, when the gating grid is approximately 90% open. At B, the sum of the voltages is $V_B = -112.2$ V which is 2.2 V below $V_a = -110$ V.

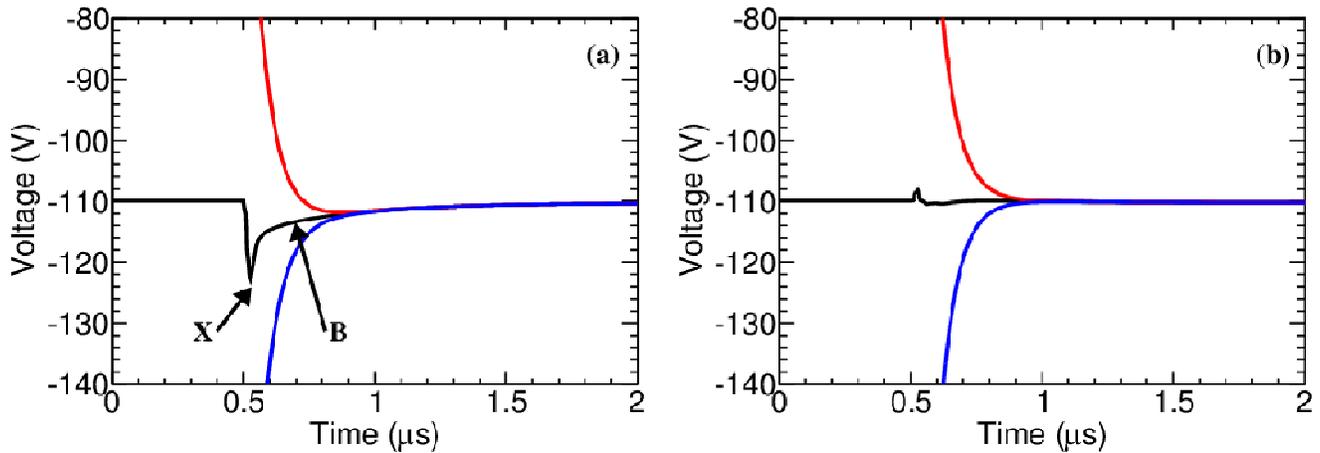

**Figure 7:** *(a) Transition of the gating grid from closed to open state in SPICE simulation, Cp=600 pF, Cn=100 pF and Rp=0.95 Ω, Rn=1.05 Ω. (b) is the same as (a) with different capacitor values Cp=600 pF, Cn=600 pF and Rp=0.95 Ω, Rn=1.05 Ω.*

To show how one can minimize this offset, we vary the values of Cn and Cp as listed in Table 2, while holding Rp and Rn constant at Rp=0.95 Ω, Rn=1.05 Ω. Shown in Figure 8 is the dependence of $V_X$ and $V_B$ as a function of (Cp−Cn). The offsets at points X and B can be shifted by increasing Cn or Cp. As expected the best configuration is when Cn is comparable to Cp. One can approximately cancel the effect of the large value of Cp=600 pF by balancing it with a comparably large value of Cn=600 pF. The result is very similar to that for the default values of Cp=Cn=100 pF and Rp=0.95 Ω, Rn=1.05 Ω. To determine the uncertainties for this calculation, 100 trials were run while allowing Cp and Cn to vary by 5%, a typical tolerance for the components used. The standard deviation of the resulting $V_X$ and $V_B$ was used for error bars. Figure 8 compares the gating grid performance with varying values of Cp and Cn. Blue squares denote where Cn is held constant at 100pF while Cp is varied, blue circles denote where Cp is held constant at 100 pF and Cn is varied, blue inverted triangles denote Cn=Cp=100pF, and red triangles denote Cn=Cp=600pF. For both Cp=Cn=600 pF and Cp=Cn=100 pF, the discharge rate is now nearly symmetric and both the $V_X$ and $V_B$ values are much smaller as shown in Table 2.

| SPICE simulation data | | | | | | |
|---|---|---|---|---|---|---|
| Cp (pF) | Cn (pF) | Cn-Cp (pF) | $V_X$ (V) | $\sigma_{VX}$ (V) | $V_B$ (V) | $\sigma_{VB}$ (V) |
| 600 | 100 | -500 | -12.07 | 0.44 | -3.29 | 0.18 |
| 500 | 100 | -400 | -10.23 | 0.44 | -2.57 | 0.15 |
| 400 | 100 | -300 | -7.91 | 0.42 | -1.89 | 0.12 |
| 300 | 100 | -200 | -5.29 | 0.38 | -1.17 | 0.09 |
| 200 | 100 | -100 | -2.28 | 0.3 | -0.44 | 0.06 |
| 100 | 100 | 0 | 2.9 | 0.15 | 0.3 | 0.04 |
| 100 | 200 | 100 | 5.47 | 0.23 | 1.03 | 0.06 |
| 100 | 300 | 200 | 8.48 | 0.42 | 1.76 | 0.09 |
| 100 | 400 | 300 | 11.15 | 0.38 | 2.5 | 0.13 |
| 100 | 500 | 400 | 13.23 | 0.39 | 3.19 | 0.17 |
| 100 | 600 | 500 | 14.95 | 0.39 | 3.92 | 0.22 |
| 600 | 600 | 0 | 2.4 | 0.58 | 0.27 | 0.22 |

**Table 2:** *List of nominal Cn, Cp values used in 100 SPICE calculations and resulting voltages at points X and B. Error for $V_X$ and $V_B$ are calculated by allowing Cn, Cp to vary by 5% in 100 trials. Rp and Rn are held constant at Rp=0.95 Ω, Rn=1.05 Ω.*

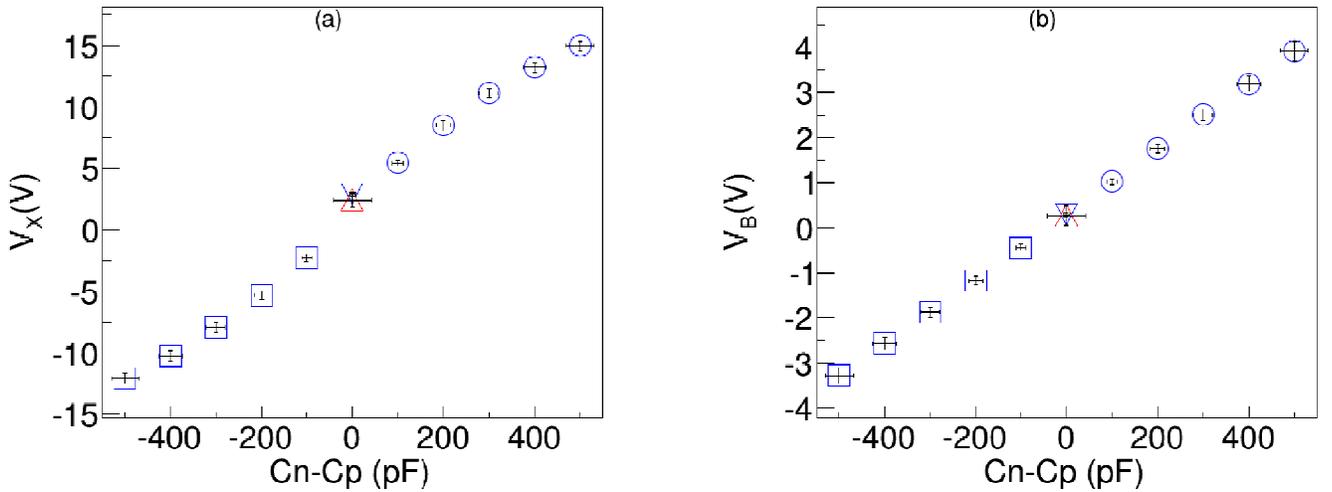

**Figure 8:** *Dependence of $V_X$ and $V_B$ values on the difference of the values of Cn and Cp. $V_X$ and $V_B$ are the voltages at the points X and B shown in Figure 7 and defined in the text. Blue squares denote where Cn is held constant at 100pF while Cp is varied, blue circles denote where Cp is held constant at 100 pF and Cn is varied, blue inverted triangles denote Cn=Cp=100pF, and red triangles denote Cn=Cp=600pF.*

IV.    **Operation Experience with SπRIT TPC**

A version of the gating grid driver circuit similar to Figure 5 has been successfully used with the SπRIT TPC during its first series of experiments at the Radioactive Ion Beam Factory (RIBF) in RIKEN, Japan in May 2016. Two commercial 4 ohm transmission lines are connected to the two sides of the gating grid. One conductor on each line carries the voltages to the positive wires and the other conductor is connected to the negative wires. Double LEMO connectors connect these transmission lines to two custom-made internal stripline transmission lines of similar design impedance that run next to the gating grid circuit boards to which the ends of the gating grid wires are attached. The internal transmission lines are connected to the gating grid on each side of the pad plane by multiple connections distributed along the gating grid. The impedance of the internal transmission lines and connections, while small, is nontrivial because the transmission line on each side has 28 connections through the gating circuit boards to the positive gating grid wires and 28 analogous connections to the negative wires; these connections are evenly dispersed along the length of the gating grid. As the driver has no magnetic components, we used a relatively short cable of approximately 1 meter between TPC and the gating grid driver to minimize reflections.

The TPC was installed inside the SAMURAI dipole magnet, which was set to a magnetic field of 0.5 T. To account for the environment of the TPC, Rp and Rn were set to the values of 3.52 ohm and 4 ohm respectively. These values were chosen empirically to minimize the induced noise on the pads when the gating grid opens. In this particular case, it was not necessary to use Cp and Cn. The Garfield calculations in Figure 4 shows that the transparency is very close to zero at $\Delta V=0.5\times(V_h-V_l)$ ~50 V. To provide a margin of safety to account for leakage from electron scatterings, we used a value of $\Delta V=75$ V in the operation of the TPC.

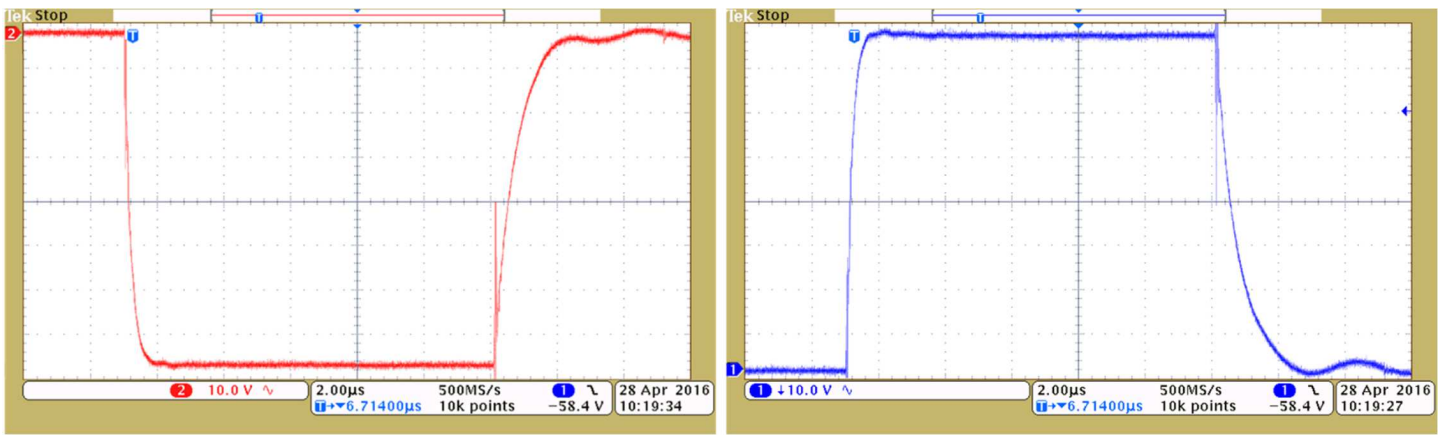

**Figure 9:** *The scope traces of the voltages measured at the $V_h$ (left panel) and $V_l$ (right panel). The TTL1 signal is provided about 3 µs from the start of the traces.*

Figure 9 shows the scope traces of the voltages measured at $V_h$ (left panel) and $V_l$ (right panel). To record these traces we ran 4m long 50 ohm LEMO cables from the Gating grid driver in the SAMURAI magnet to an oscilloscope situated in a region of low magnetic field. The scope allowed measurements of these large signals only when terminated at high input impedance, which introduced the reflections at 40 ns intervals observed on the traces. The traces start 3 $\mu$s prior to the TTL1 signal being provided to open the gating grid. The TTL1 signal persists until the gating grid is closed by switching off the TTL1 signal and providing the TTL2 signal 11 $\mu$s after opening. The average voltage in this test was set to be $V_a=-170.83$ V, which is ~60 V more negative than the simulations. This difference in $V_a$ does not materially influence the performance of the gating grid. At the beginning of the scope traces, both $V_h$ and $V_l$ have their nominal values of −95.6 V and −246.06 V ($\Delta V$=75.2 V) and the gating grid is closed. When the driver receives the TTL1 signal, the alternate gating grid wires are shorted to the average voltage of $V_a=-170.83$ V as discussed in Section III. The opening time of the gating grid is indicated by the rise time of the scope trace after the TTL1 signal is provided. The sum of the traces for $V_h$ and $V_l$ provides the time dependence of average voltage $V_a$ shown in the left panel and the difference between $V_h$ and $V_l$ provides values for $\Delta V$ shown in the right panel of Figure 10. The reflections of the signals due to the high input impedance of the scope inputs obscures somewhat the overall exponential decay of $\Delta V$ at first, but it is clear at later times that $\Delta V$ drops to less than 10 V within 350 ns, with $\Delta V \leq 10$ V corresponding to at least 90% electron transparency according to the Garfield calculations shown in Figure 4.

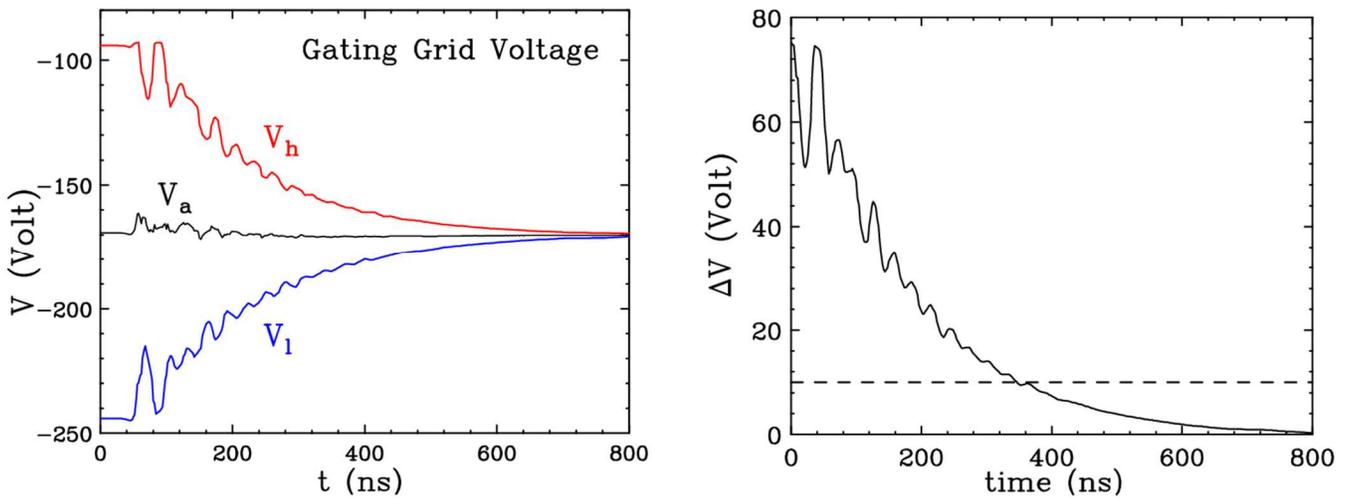

**Figure 10:** *Digitized values of the scope traces in Figure 9. a. (left panel) $V_h$, $V_l$, and $V_a$ as a function of time. The traces begin 50 ns before the TTL1 signal. b. (right panel) the time dependence of $\Delta V$.*

The first series of experiments with S$\pi$RIT TPC are currently being analyzed. Preliminary results including those presented in Figures 9 and 10 show that the gating grid with the driver worked successfully, with only a small pickup correction that can be completely removed from the data at t $\geq$600 ns. The performance of the gating grid with analyzed experimental data will be described in detail in an upcoming article.

V. **Summary and Conclusion**

A new gating grid driver has been designed for use with Time Projection Chambers and other similar devices that require a gating grid that operates in a bipolar mode with different electrostatic potentials on alternating wires. To open such gating grids, the driver shorts the alternate wires to a common voltage **V$_a$**. Later, this driver closes the grid by restoring the voltage differences in the adjacent wires. We have used the SPICE circuit analysis program to analyze the properties of the circuit. It opens the gating grid in 0.20 $\mu$s, minimizing the lost drift length associated with this opening time. The circuit consists of 2 pairs of N- and P- MOSFET switches and includes two adjustable capacitors and resistors that can be used to adjust the opening time, and shift the balance of positive and negative charge for individual TPC. A gating grid driver based on the new design allows the gating grid to open within 350 ns in the first series of experiments using the S$\pi$RIT TPC.

## Acknowledgements

This work is supported by the U.S. Department of Energy under Grant Nos. DE-SC0004835 and National Science Foundation Grant No. PHY 1102511.